\documentclass[a4paper,twoside]{article}

\usepackage{epsfig}
\usepackage{subcaption}
\usepackage{calc}
\usepackage{amssymb}
\usepackage{amstext}
\usepackage{amsmath}
\usepackage{amsthm}
\usepackage{multicol}
\usepackage{pslatex}
\usepackage{apalike}
\usepackage{xcolor}  
\usepackage{color}  
\usepackage{algorithm2e}
\usepackage{url}
\usepackage{tikz}
\usepackage{graphicx}
\usepackage{pgfplots}
\usepackage{multirow}%
\usepackage{float}
\usepackage[hidelinks]{hyperref}
\usepackage{caption}
\usepackage{soul}
\usepackage{booktabs}

\usepackage{fancyhdr}

\usepackage[bottom]{footmisc}
\usepackage{SCITEPRESS}     

\fancypagestyle{specialfooter}{%
  \fancyhf{}
  
  \fancyfoot[L]{\footnotesize%
  \textcopyright{} The Authors. This version did \textbf{not} undergo peer-review: please refer to the version of record below.\\
  In \emph{Proceedings of the 13th International Conference on Pattern Recognition Applications and Methods (ICPRAM 2024)}, pp. 995--1000.\\
  The final Open Access publication is available at ScitePress via \textcolor{blue}{\url{https://doi.org/10.5220/0012608600003654}}
  }
}

\begin{document}

\thispagestyle{specialfooter}

\title{Detecting Brain Tumors through Multimodal Neural Networks}

\author{\authorname{Antonio Curci\sup{1}\orcidAuthor{0000-0001-6863-872X}, Andrea Esposito\sup{1}\orcidAuthor{0000-0002-9536-3087}}
\affiliation{\sup{1}Department of Computer Science, University of Bari Aldo Moro, Via E. Orabona 4, 70125 Bari, Italy}
\email{\{antonio.curci, andrea.esposito\}@uniba.it}
}


\abstract{Tumors can manifest in various forms and in different areas of the human body. Brain tumors are specifically hard to diagnose and treat because of the complexity of the organ in which they develop. Detecting them in time can lower the chances of death and facilitate the therapy process for patients. The use of Artificial Intelligence (AI) and, more specifically, deep learning, has the potential to significantly reduce costs in terms of time and resources for the discovery and identification of tumors from images obtained through imaging techniques. This research work aims to assess the performance of a multimodal model for the classification of Magnetic Resonance Imaging (MRI) scans processed as grayscale images. The results are promising, and in line with similar works, as the model reaches an accuracy of around 99\%. We also highlight the need for explainability and transparency to ensure human control and safety.}

\onecolumn \maketitle \normalsize \setcounter{footnote}{0} \vfill

\section{\uppercase{Introduction}}
\label{sec:introduction}
Brain tumors refer to a heterogeneous group of tumors arising from cells within the Central Nervous System (CNS) \cite{WHOClassificationofTumoursEditorialBoard2022WHO}. These tumors can manifest in various forms, ranging from benign to malignant, and may originate within the brain tissue or spread from other parts of the body through metastasis \cite{Lapointe2018Primary}. In this regard, it is crucial to underline that tumors that spread in brains are incredibly complex to treat because of the extreme delicacy that the organ in question is characterized by.

Brain tumors can rise several symptoms in individuals who suffer from them, such as strong and recurring headaches, nausea, altered mental status, papilledema, and seizures; the implications of these symptoms in individuals can worsen over time if the tumor is not detected in time, resulting, eventually, in death \cite{Alentorn2016Presenting}. This implies that the prompt detection, diagnosis, and removal of tumors must be supported by proper tools and techniques to assist professionals and increase their efficiency when performing these tasks. Therefore, there is the need for ools and instruments featuring the newest technologies that can support and facilitate this process for physicians \cite{McFaline-Figueroa2018Brain}.

The aid of technology, more specifically Artificial Intelligence (AI), can provide significant advantages concerning the precision, speed, and overall efficacy of detecting these tumors, thereby improving therapy outcomes and quality of life \cite{Ranjbarzadeh2023Brain}. In fact, the landscape of AI models for the detection of brain tumors is vivid \cite{Anaya-Isaza2023Optimizing,Vermeulen2023Ultrafast,Huang2022Artificial,Ranjbarzadeh2023Brain}.

Traditionally, brain tumors are diagnosed by using imaging techniques, such as \emph{Magnetic Resonance Imaging (MRI)}, \emph{Computed Tomography (CT)}, or \emph{Positron Emission Tomography (PET)}, which are incredibly useful and effective. However, the integration of AI in this context can further improve and enhance their outputs and maximize efficiency \cite{Villanueva-Meyer2017Current}. Recent research has focused on using machine learning and deep learning techniques for brain tumor classification, segmentation, and feature extraction, as well as developing AI tools to assist neurosurgeons during treatment \cite{Vermeulen2023Ultrafast,Huang2022Artificial}.

The current scenario of the application of Neural Networks employed in the field of medicine and in brain tumor detection encompasses various models and techniques, and still represents a very challenging issue.
For instance, Mohesen et al.~use Deep Neural Networks (DNN), combined with Principal Component Analysis (PCA), and Discrete Wavelet Transform, achieving a good accuracy, around 97\% \cite{Mohsen201868}. 
Pei et al., instead, used 3D Convolutional Neural Networks (CNN), reaching a training accuracy of around 81\% and a validation accuracy of around 75\% \cite{Pei2020}. In addition, Nayak et al.~developed another CNN as a variant of Efficient DenseNets with dense and drop-out layers, obtaining an accuracy close to 99\% \cite{Nayak2022}.
The employment of these models in classification tasks in medicine can be significantly useful. At the same time, it remains crucial for professionals to maintain control and be able to check the output of these instruments to have the final say over the model's predictions. 
The employment of multimodal models, instead, is still under development and research in the literature. It is possible to find cases in which these models are built with Multi-Layer Perceptrons (MLP) or with DenseNets for 3D images image classification, in which researchers could not achieve high-performance rates \cite{Ma2020,Latif2017}.
Different modalities provide different types of information. Images can visual information about the tumor's location, size, and external characteristics, while tabular data can include insights about other aspects and peculiarities either highlighted by the physician or numerical data extracted from the images themselves. Combining these modalities can improve the AI model when it comes to learning how to discriminate between tumor and non-tumor cases. Multimodal AI can also provide a more comprehensive decision support system for healthcare professionals, leading to better clinical decision-making and treatment planning \cite{Soenksen2022,Yang2016}.

This research work aims at creating and employing a multi-modal model to classify brain images as healthy or ill (i.e., containing a tumor) and proposing an approach towards stronger explainability and transparency to increase physicians' trust levels when using AI in medicine. The model in question was built through a Densely Connected Convolutional Network (DenseNet) and it was trained over a labeled dataset composed of tabular data and 2D brain tumor images.

\begin{figure}[b]
    \centering
    \begin{subfigure}[b]{.45\linewidth}
        \centering
        \includegraphics[width=\linewidth]{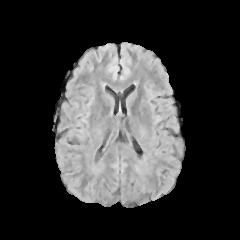}
        \caption{Healthy scan}
        \label{fig:images0}
    \end{subfigure}
    \vspace{0.5cm}
    
    \begin{subfigure}[t]{.45\linewidth}
        \centering
        \includegraphics[width=\linewidth]{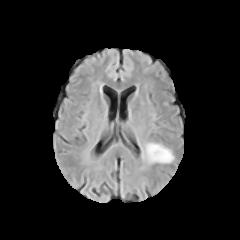}
        \caption{Scan presenting a tumor}
        \label{fig:images1}
    \end{subfigure}
    ~
    \begin{subfigure}[t]{.45\linewidth}
        \centering
        \includegraphics[width=\linewidth]{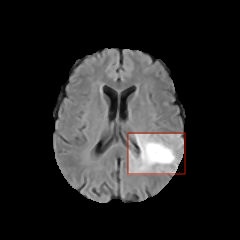}
        \caption{Highlighted lesion of the ill brain}
        \label{fig:images1zoom}
    \end{subfigure}
    \caption{Examples of MRI scans available in the dataset}
    \label{fig:data-example}
\end{figure}

This paper is organized as follows: \autoref{sec:materials} encompasses all the materials used during this study, defining the dataset, its provenance, and the distribution of the classes; \autoref{sec:methods} explores the model, its structure, and the parameters set for the experiment. In \autoref{sec:results}, we describe the tools used to carry out the experiment and we analyze the results; \autoref{sec:conclusions}, instead, provides an overview of the research work, its outcomes, and the future directions that we intend to undertake for this project, highlighting the need for explainability and control.

\begin{figure*}[h!]
    \centering
    \includegraphics[width=.9\textwidth]{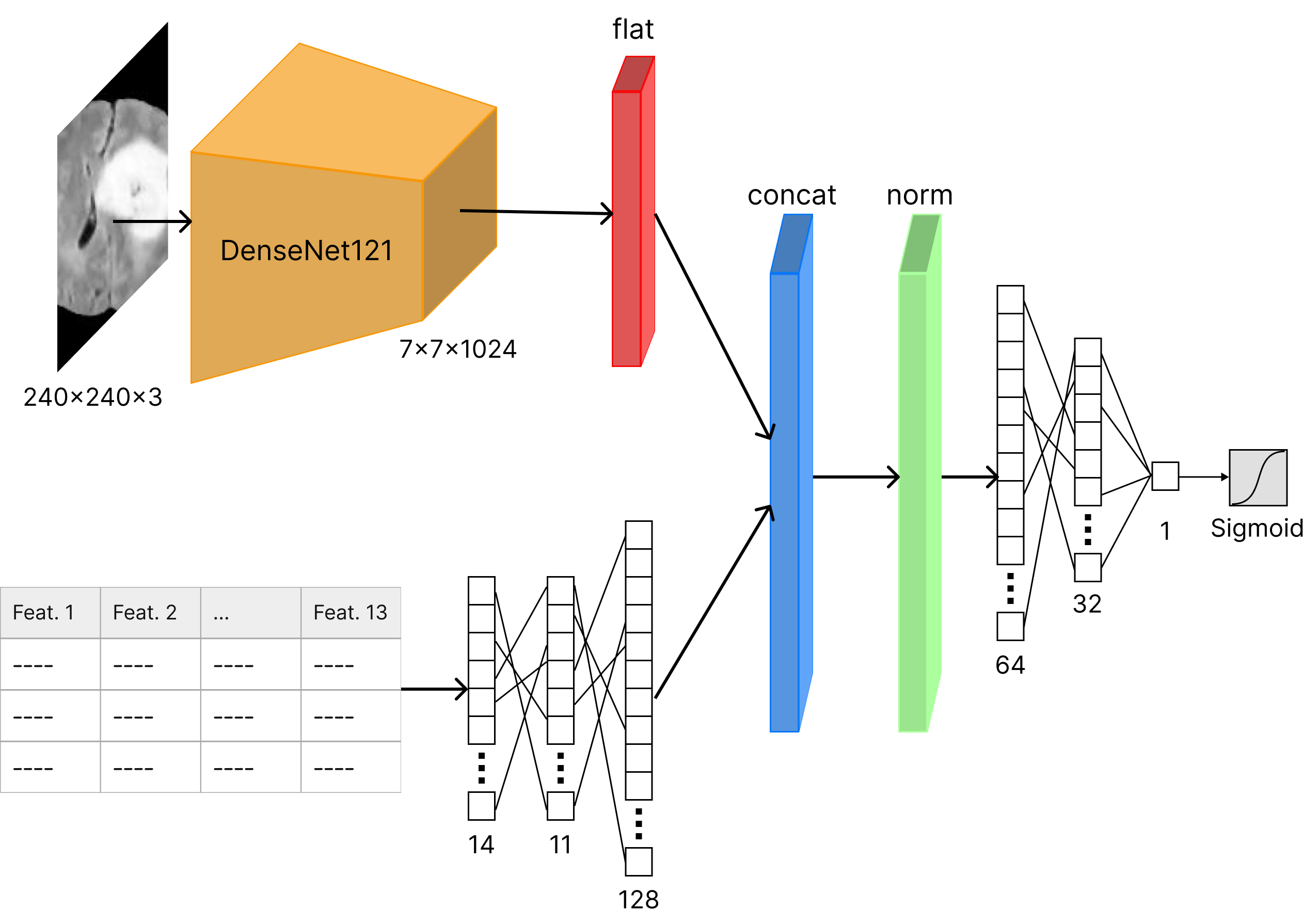}
    \caption{Architecture of the multi-modal deep neural network}
    \label{fig:plotmodel}
\end{figure*}

\section{\uppercase{Materials}}
\label{sec:materials}
This research work was conducted using a dataset derived from the BRATS 2015 challenge \cite{Menze2015Multimodal}, freely available on \textit{Kaggle.com} \cite{JakeshBohajuBrain}. The dataset comprises 3762 instances. Each instance consists in a $240\times 240$ three-channel MRI scans of the brain, and in a set of 13 numeric features (with an additional feature that allows to identify the scan associated with the numeric values).
The dataset is fully labeled. The labels are binary and mutually exclusive: a value of ``0'' represents the absence of a tumor (in the following, we will refer to this class as ``healthy''); a value of ``1'' indicates the presence of a tumor (in the following, we will refer to this class as ``ill''). The tabular data has 13 features of first- an second-order; they were extracted by the authors of the dataset from the images, which are the processed output of MRI scans. The first-order features are \textit{Mean}, \textit{Variance},\textit{Standard Deviation}, \textit{Skewness}, \textit{Kurtosis}, while those of second-order are \textit{Entropy}, \textit{Contrast}, \textit{Entropy}, \textit{Energy},  \textit{Dissimilarity}, \textit{Correlation}, \textit{Coarseness}, \textit{ASM (Angular second moment)}, \textit{Homogeneity}.

The dataset is slightly unbalanced, with 2079 instances labeled as healthy and 1683 labeled as ill. To avoid the potential introduction of artifacts or unrealistic samples using data augmentation \cite{Chlap2021Review}, the class-imbalance problem was solved by dropping randomly selected instances from the ``healthy'' class. The numeric features of the dataset have been standardized in order to have mean $\mu=0$ and variance $\sigma^2=1$. The dataset has no missing values, making it unnecessary to perform any additional pre-processing.

\autoref{fig:data-example} provides examples of images labeled as healthy and ill. More specifically, \autoref{fig:images1} is the image of the scan of a brain containing a tumor, which is found in its lower-right part as a white area that stands out from the rest of the organ; the latter is pointed in  \autoref{fig:images1zoom} in the highlighted red rectangle.

%

\section{\uppercase{Methods}}
\label{sec:methods}

The model used for this research work is a multi-modal neural network. The model architecture, depicted in \autoref{fig:plotmodel}, is composed of two heads (one for each type of input data). The first head is responsible for the feature extraction from the MRI scans: it consists in a DenseNet121 network \cite{Huang2018Densely} with input size $240\times 240\times 3$ and output size $7\times7\times1024$, that is then flattened. The second head, responsible for the encoding of the tabular data, consists in a simple fully-connected neural network, using the Rectivied Linear Unit (ReLU) activation function. The outputs of the two heads are then concatenated and normalized. The resulting vector is then provided as input to an additional fully-connected neural network (also using the ReLu activation function), which terminates in two SoftMax-activated neurons that provide the final prediction.
The model is shown in \autoref{fig:plotmodel}.


\section{\uppercase{Results}}
\label{sec:results}

\begin{table*}[h!]
    \centering
    \begin{tabular}{@{}ccccccc@{}}
        \toprule
        CV Fold & Accuracy &  AUC &    Loss & Precision & Recall & F1-Score\\
        \midrule
1       &     0.99 & 0.99 &    0.18 &      0.99 &   0.98 &     0.99 \\
2       &     0.97 & 0.97 &     1.5 &      0.99 &   0.95 &     0.97 \\
3       &     0.99 & 0.99 & 5.6e-05 &      0.99 &   0.99 &     0.99 \\
4       &     0.98 & 0.98 &     1.3 &      0.98 &   0.98 &     0.98 \\
5       &     0.97 & 0.98 &    0.67 &      0.95 &   0.99 &     0.97 \\
6       &     0.99 & 0.99 &    0.84 &      0.99 &   0.99 &     0.99 \\
7       &     0.99 & 0.99 &    0.72 &      0.99 &   0.99 &     0.99 \\
8       &     0.98 & 0.98 &     2.9 &      0.99 &   0.96 &     0.98 \\
9       &     0.99 & 0.99 &    0.22 &      0.99 &   0.99 &     0.99 \\
10      &     0.99 & 0.99 &       0 &      0.99 &   0.99 &     0.99 \\
		\midrule
Avg.	&     0.99 & 0.99  &    0.83 &     0.99 &   0.98 &    0.98 \\
        \bottomrule
    \end{tabular}
    \caption{Results of the cross validation}
    \label{tab:cv-res}
\end{table*}

The experiment was performed using an Apple Silicon M2 Pro chip with an integrated 16-core GPU, using the TensorFlow library. 

To evaluate the proposed method, a \emph{stratified} 10-fold cross-validation was used (i.e., each fold contained roughly the same proportion of the two class labels). For the training phase, we used binary cross-entropy as the loss function, defined in Equation \ref{eq:sparsecatcrossentropy}, where $y_i$ is the ground truth label, while $p_i$ is the model output for an individual observation.

\begin{equation}
    \label{eq:sparsecatcrossentropy}
    \mathcal{H}(\mathbf{y},\mathbf{p})=
    \frac{1}{N}
    \sum\limits_{i=1}^N-(y_i\log(p_i)+(1- y_i)\log(1- p_i))
\end{equation}

Cross-entropy was minimized using the Adam optimizer, with a static learning rate of $10^{-3}$ and a batch size of $32$. The maximum number of epochs was set to $10^2$, with an early stopping criterion based on the validation loss with a minimum delta of $10^{-4}$ and a patience of $5$ epochs.

As performance metrics, we opted for the most commonly used metric in classification problems:
\begin{itemize}
    \item Accuracy: defined as the proportion of the correctly classified samples (both positives and negatives) in the selected population.
    \item Recall: which refers to the proportion of diseased subjects who have been classified as ill;
    \item Precision: that is the proportion of the correctly classified samples among all ill-classified samples;
    \item F1-Score: that is the harmonic mean between the precision and recall;
    \item Area Under ROC-Curve (AUC): that indicates the probability that, given a healthy and an ill sample, the classifier is able to correctly distinguish them.
\end{itemize}



\begin{figure}[b]
    \centering
    \includegraphics[width=\linewidth]{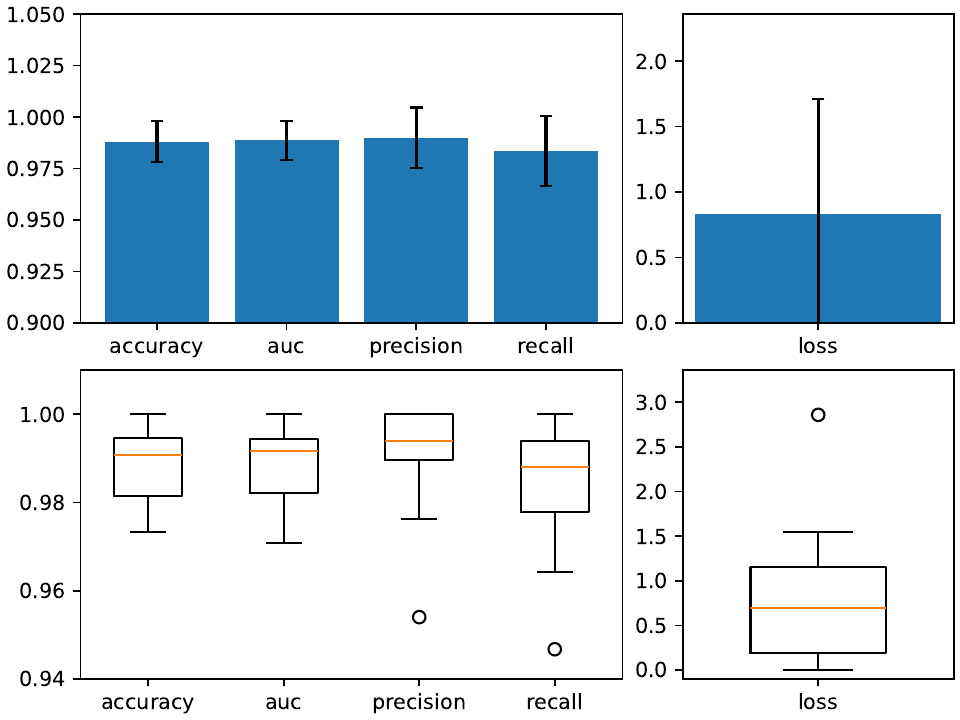}
    \caption{Results of the cross-validation}
    \label{fig:res}
\end{figure}

The training phase on the 10 folds exhibited quite good performances, shown in \autoref{tab:cv-res}; each fold generated accuracy rates higher than 97\%, with an average of 98.80\%. The average values for all metrics are available in \autoref{tab:cv-res}. The loss has values less than 1.5, as shown in \autoref{fig:plotmodel}. The only exception is the eighth fold, which has a loss value close to 2.9: further inspection is needed to uncover the reasons for this sudden peak.

\section{\uppercase{Conclusions and Future Works}}
\label{sec:conclusions}
In this article, we explore the use of multi-modal DenseNets for brain tumor images classification. The presented model is useful when dealing with data of different types with intrinsically different representations, in this case, tabular data and images. The multi-modal Deep Neural Network created and exploited in this case study provides promising results for classifying brain tumor images, achieving an average accuracy of 98\%. The results are on par with other techniques found in the literature \cite{Nayak2022,Mohsen201868}.

Although the dataset used in this work was also used by other researchers in the community, a multi-modal model was never chosen as the approach to undertake to perform a classification task. 
Arora et al created a model consisting of a Convolutional Neural Network (CNN) that reached a 90\% accuracy using a VGG16 Neural Network \cite{Arora2021,Simonyan2015}. Herm et al, instead, employed a CNN with 13 layers, obtaining an accuracy of around 89\% \cite{Herm2023}.
Another research work was performed on this dataset by Morris L., whose model achieved 87\% of accuracy by using a Deep Neural Network with MobileNetV2 \cite{Morris2022,Sandler2019}.
It emerges that the work presented in this article provides a starting ground for future research, exploring how the exploitation of different types of data can be for the classification of brain tumor images. 

An initial aspect that needs further exploring is the model generalizability: future work may delve in testing the model with additional parameters and/or another dataset with the same structure and belonging to the same medical domain, to observe its behavior and efficacy in different settings.
Moreover, future work involves also the comparison of the performance of this multi-modal model with standard classifiers, meaning models that are trained merely on tabular data or on images. The objective is to determine how the characteristics of the model proposed in this work can be beneficial to the medical field with respect to a more traditional approach.

In addition, explainability and transparency are needed to provide users (i.e., physicians) with more efficient instruments to understand and comprehend the outputs it provides. As neural networks' outputs are usually obscure to users without expertise in computer science and, specifically, in AI, explainability has the potential of demystifying the process that lies behind the final predictions and output of models. Moreover, it is crucial for physicians to fully understand the reasons why an AI systems provided a specific outcome \cite{Combi2022Manifesto}, as this ensure human control. In fact, from an ethical point of view, the responsibility that physicians undertake when making decisions about the health state of their patients cannot depend merely on algorithms that they do not comprehend properly.
Explainability plays an important role for physicians because it allows to check and keep track of which features were relevant for the prediction outputted by the AI model and detecting potential mistakes that can be corrected thanks to their expertise. The motivation behind this lies in the fact that AI systems are never perfectly accurate, thus, the clinical revision process has to be carried out precisely and meticulously by professionals, implying that having complete and blind trust is not feasible for legal reasons, too \cite{Amann2020}.
It emerges that the goal is to approach a \emph{symbiotic} relationship between AI and humans. The use of AI in medicine, especially Neural Networks, can be beneficial both diagnostically and to foster and guide future research (e.g., through machine teaching \cite{Selvaraju2016GradCAM}).

The multi-modal neural network presented in this article provides an interesting proving ground, to evaluate the balance between accuracy, model complexity, and explainability in a challenging high-risk domain.

\section*{\uppercase{Acknowledgements}}

The research of Antonio Curci is supported by the co-funding of the European Union - Next Generation EU: NRRP Initiative, Mission 4, Component 2, Investment 1.3 -- Partnerships extended to universities, research centers, companies, and research D.D. MUR n. 341 del 15.03.2022 -- Next Generation EU (PE0000013 -- ``Future Artificial Intelligence Research -- FAIR'' - CUP: H97G22000210007).

The research of Andrea Esposito is funded by a Ph.D.~fellowship within the framework of the Italian ``D.M.~n.~352, April 9, 2022'' - under the National Recovery and Resilience Plan, Mission 4, Component 2, Investment 3.3 - Ph.D. Project ``Human-Centered Artificial Intelligence (HCAI) techniques for supporting end users interacting with AI systems'', co-supported by ``Eusoft S.r.l.'' (CUP H91I22000410007).

\bibliographystyle{apalike}
{\small
\bibliography{BrainTumorDenseNet}}
\end{document}